# Modification to the Fresnel formulas for amplified or attenuated internal reflection

**Takeyuki Kobayashi**

Department of Opto-Electronic System Engineering, Chitose Institute of Science and

Technology, 758-65 Bibi, Chitose, Hokkaido 066-8655, Japan

E-mail: ta-koba@photon.chitose.ac.jp

The present study addresses the question of total internal reflection of a plane wave from an amplifying or attenuating medium. Inspection on the expressions for the modal gain or loss of an asymmetric planar waveguide having an amplifying or attenuating cladding has led us to the coefficients for amplified or attenuated internal reflection. The derived formulas suggest an interpretation that the refracted inhomogeneous plane wave undergoes amplification or attenuation while it travels a distance equal to the Goos-Hänchen shift along the boundary plane before going back into the high-index medium. Furthermore, the evanescent wave in the low-index medium is discussed. Unlike the transparent case, the Poynting vector is shown to have a nonvanishing time-averaged component perpendicular to the boundary plane.

# I. Introduction

Since its conception [1], there have been many reports on the investigation of optical amplification by means of total internal reflection from an amplifying low-index medium. This idea, which goes by a few different names such as evanescent-field amplification, evanescent gain, amplified total internal reflection, and enhanced internal reflection, has been implemented experimentally by use of various structures [2-7], where luminescent molecules in organic matrices were used most likely because of their large transition cross-sections, which make it easy to observe amplification and lasing. Most notably, in [4], an intensity increase by a factor of 1000 was reported.

As for its theoretical understanding, however, the mechanism for the phenomenon has yet to be established. Analytical and numerical studies have been undertaken [8-10]. Based on the theory developed in Ref. [8], the upper limit of reflectance was estimated to be 5.83 in Ref. [4]. In Ref. [9], it was shown that the reflectance for amplification is given by the inverse of the reflectance for absorption if one reverses the sign of the extinction coefficient. In Refs. [11, 12], an exponential excitation gradient along the boundary was assumed to facilitate optical gain. The experimental demonstration of a reflectance of 1000 [4] was supported by work in Ref. [10]. Some resorted to numerical calculation to show that a greater-than-unity reflection coefficient is possible [13, 14]. A unique approach was taken in Ref. [15], where attention was paid to the problem of causality in the theory of classical electrodynamics to conclude that, with a suitably designed gain medium, it is possible to realise evanescent-field amplification. It has even been argued that a reflection coefficient exceeding unity is unfeasible [16]. What most of these past theoretical studies have in common is that they sought a physical picture in which intensity growth takes place in the direction perpendicular to the interface between the transparent high-index medium and the low-index gain medium. None of the results thus obtained, however, has been applied to other physical situations that involve amplified internal reflection to see if they make any physical sense.

Among such situations, the most-often encountered and experimented are waveguides that exhibit gain or loss. The semiconductor heterostructure, for example, consists of a waveguiding layer with lower-index cladding regions, all of which exhibit either gain or loss. Solving Maxwell's equations for modal gain or loss with suitable boundary conditions yields solutions whose spatial intensity evolution in the $x$ direction is of the form: $\exp(\Gamma\gamma x)$, where $\Gamma$ is the fractional overlap between the guided-mode power flow and the gain or loss medium and $\gamma$ is the material gain or loss coefficient [12].

In the present study, instead of inserting a complex refractive index in place of the real one in the Fresnel formulas, we take a different route: starting with the expressions for modal gain or loss like the one in the foregoing paragraph, which may as well be viewed as the result of successive multiple reflections of plane waves, we work out formulas for a single total reflection from an amplifying or attenuating lower-index medium. The problem of finding the reflection coefficients thus becomes straightforward.

After the introduction, the reflection coefficients are derived in Sec. II. Sec. III discusses the evanescent field, which is shown to give a nonzero Poynting-vector component perpendicular to the boundary plane. Sec. IV concludes the study.

## II. Derivation of the reflection coefficients

Total internal reflection is accompanied by phase changes, which are given by the Fresnel reflection coefficients. For an infinite plane wave incident at an angle $\alpha$ on a plane boundary separating two transparent media having indices of refraction $n_h$ and $n_l$ ($n_h > n_l$), we have

$$r_{s0} = \exp(-j\Phi_s), \tag{1}$$

where $\Phi_s = 2\arctan[(n_h^2\sin^2\alpha - n_l^2)^{1/2}/n_h \cos \alpha]$ for $s$-polarized waves, and

$$r_{p0} = \exp(-j\Phi_p), \tag{2}$$

where $\Phi_p = 2\arctan\{(n_h/n_l)^2[(n_h^2\sin^2\alpha - n_l^2)^{1/2}/n_h \cos \alpha]\}$ for $p$-polarized waves [13]. Here the $s$-component means the electric field perpendicular to the plane of incidence, and the $p$-component the electric field parallel to the plane of incidence. In what follows, we assume that, if the coefficient that governs the amplitude change is written as either $\rho_s$ or $\rho_p$, depending on the polarization, the reflection coefficient with the phase and amplitude changes altogether is obtained by the product: $r_s = r_{s0}\, \rho_s$ for $s$-polarization or $r_p = r_{p0}\, \rho_p$ for $p$-polarization.

Here an asymmetrical planar waveguide is assumed (figure 1), whose core region of thickness $T$ and index of $n_h$ is sandwiched by the amplifying or attenuating cladding region of index $n_l - j\kappa$ and transparent substrate of index $n_s$. The imaginary part of the index $\kappa$ is sometimes called the extinction coefficient and accounts for attenuation for $\kappa > 0$ in the present study. Also we assume that $n_l \gg |\kappa|$ so the mode profile remains virtually unaffected by the presence of gain or loss. Even for analysing semiconductor amplifiers whose gain is as large as or larger than polymer gain media, this is a reasonable assumption, which holds for all cases of practical interest, because the ratio of the imaginary to the real part of the index is so small [19]. Just to take a look at how small the ratio could be with dye-doped polymers, we take, as an example, a doped polymer gain medium, which exhibited a modal gain coefficient of $\gamma = 106$ cm$^{-1}$ at 0.5 μm, the largest value measured to date in the nanosecond region [20]. The value translates to a material gain of ~130 cm$^{-1}$ or $\kappa \sim -5 \times 10^{-4}$. With $n_l = 1.58$, we get the ratio $|\kappa|/n_l \sim 3 \times 10^{-5}$.

We imagine a guided mode, either TE or TM, with an incident angle of $\alpha$. Discussion on the phase evolution is found in literature in relation to the waveguiding conditions (see, for example, [18], [21] or [22]) and is of no interest to us for the moment. While traversing a distance $L$ in the $x$ direction, the guided mode experiences gain or loss as a result of successive multiple total reflections at the boundary between the core and the amplifying or attenuating cladding. First we consider TE modes, which are regarded as $s$-polarized plane waves superposed in phase. The modal gain or loss $G_s$ is expressed as

$$G_s = \exp(-2\Gamma_s^l k_0 \kappa L), \qquad (3)$$

where $\Gamma_s^l$ is the overlap factor in the cladding region, and $k_0$ is the wavenumber in vacuum. The extinction coefficient $\kappa$ is related to the material gain ($\kappa < 0$) or loss ($\kappa > 0$) coefficient by $\gamma = -2k_0\kappa$. We see from figure. 1 that the total number of reflections at the cladding-core boundary for a TE mode is given by $N = L/\Delta L_s$. Here $\Delta L_s$ is the distance between two successive reflections at the core-cladding interface: $\Delta L_s = 2(d_s^l + T + d_s^s) \tan \alpha$. The penetration depths $d_s^l$ and $d_s^s$ in the cladding and substrate regions, defined as the distance from the boundary plane where the amplitude drops to $1/e$ of the value at the boundary, can be expressed as

$$d_s^l = k_0^{-1}(n_h^2 \sin^2\alpha - n_l^2)^{-1/2} \qquad (4)$$

and

$$d_s^s = k_0^{-1}(n_h^2 \sin^2\alpha - n_s^2)^{-1/2}, \qquad (5)$$

where $k_0$ is the free-space wavenumber.

The overall intensity change after traversing the waveguide of length $L$ is a result of $N$ reflections at the core-cladding boundary, each of which gives an intensity change $\rho_s^2$:

$$\exp(-2\Gamma_s^l k_0 \kappa L) = \rho_s^{2N}, \qquad (6)$$

from which we obtain

$$\rho_s = \exp(-\Gamma_s^l k_0 \kappa \Delta L). \qquad (7)$$

Here the overlap factor $\Gamma_s^l$ can be expressed explicitly in terms of the refractive indices $n_l$, $n_h$, and $n_s$ and core layer thickness $T$[1]

$$\Gamma_s^l = d_s^l n_l^2 \cos^2\alpha (d_s^l + T + d_s^s)^{-1}(n_h^2 - n_s^2)^{-1}. \qquad (8)$$

We introduce a new quantity defined as $\delta_s = \Gamma_s^l \Delta L_s$, which, with the help of $\Delta L_s = 2(d_s^l + T + d_s^s) \tan \alpha$, becomes

$$\delta_s = 2\cos\alpha \sin\alpha\, [1 - (n_l/n_h)^2]^{-1} d_s^l. \tag{9}$$

Equation (9) turns out to be what is known as the Goos-Hänchen shift for *s*-polarized waves. The expression, derived by use of the energy conservation, holds at any incident angle, not just in the vicinity of the critical angle [23]. The Goos-Hänchen shift being essentially a concept associated with optical beams of finite cross-sections, the central part of the incident wave is treated as a perfect plane wave in first approximation in the derivation of equation (9). Indeed, in the analysis of planar waveguides, viewing the guided modes as a superposition of perfect plane waves, we are able to correctly estimate the accumulated phase shift resulting from multiple total reflections at the interfaces of the core layer [18, 21, 22].

Using the Goos-Hänchen shift $\delta_s$, we find for the amplitude change: $\rho_s = \exp(-k_0 \kappa \delta_s)$. Including the phase factor, we have arrived at

$$r_s = r_{s0}\, \rho_s = \exp(-j\Phi_s)\exp(-k_0 \kappa \delta_s). \tag{10}$$

The very same line of reasoning for TM modes with a little more tedious algebraic manipulation leads us to the reflection coefficient for *p*-polarized waves. Using

$$\varGamma_p^l = d_p^l\, n_l^2 \cos^2\alpha\, (d_p^l + T + d_p^s)^{-1}(n_h^2 - n_s^2)^{-1} \tag{11}$$

and $\varDelta L_p = 2(d_p^l + T + d_p^s)\tan\alpha$, where $d_p^l$ and $d_p^s$ are given by the following

$$d_p^l = n_l^2(n_h^2 \sin^2\alpha - n_l^2 \cos^2\alpha)^{-1}\, d_s^l \tag{12}$$

and

$$d_p^s = n_s^2(n_h^2 \sin^2\alpha - n_l^2 \cos^2\alpha)^{-1}\, d_s^s, \tag{13}$$

we find

$$r_p = r_{p0}\, \rho_p = \exp(-j\Phi_p)\, \exp(-k_0 \kappa \delta_p), \tag{14}$$

where the Goos-Hänchen shift for *p*-polarized waves is written as

$$\delta_p = 2(n_l/n_h)^2 \cos\alpha \sin\alpha\, [(n_l/n_h)^4 \cos^2\alpha + \sin^2\alpha - (n_l/n_h)^2]^{-1} d_s^l. \tag{15}$$

We observe that, for $\kappa < 0$, a reflection coefficient exceeding unity naturally comes about. A given point on the wavefront, upon its incidence, traverses a path $\delta_s$ or $\delta_p$, depending on the polarization, along the boundary plane and then goes back into the high-index medium [23]. In [23], the author also emphasizes that the evanescent wave is a travelling wave, not a standing one, contrary to the oft-stated notion in literature. If this is the case, equations (10) and (14) suggest that the stimulated transitions responsible for evanescent gain or loss in the low-index medium are induced by the inhomogeneous plane waves travelling a distance $\delta_s$ or $\delta_p$ along the boundary.

The past theoretical works that support the existence of such evanescent gain essentially assume that the amplitude growth by stimulated emission somehow occurs in the direction perpendicular to the boundary plane. This, however, presents a physical picture incompatible with the textbook knowledge of stimulated transitions in that the amplification occurs perpendicular to the time-averaged Poynting vector of the evanescent wave travelling parallel to the boundary [18]. Note that the formulas proposed here, equations (10) and (14), facilitate the theory for its mechanism outlined in the foregoing paragraph, allowing us to reconcile the occurrence of such phenomenon with the conventional idea of optical amplification by stimulated emission.

We notice that the proposed formulas go over to amplification from attenuation or vice versa by merely reversing the sign of the imaginary part of the index, as should be the case with the theory of classical electrodynamics, and also that they reduce, as expected, to the original Fresnel reflection coefficients as the imaginary part $\kappa$ approaches zero.

Now we are in the position to carry out an order-of-magnitude consideration on amplified internal reflection. Noting that $\delta_s$ is known to be on the order of $k_0^{-1}$ [18], we have $k_0\delta_s \sim 1$. If we use $\kappa \sim -5 \times 10^{-4}$ quoted above, we find $\rho_s = \exp(-k_0\kappa\delta_s) \sim 1 - k_0\kappa\delta_s \sim 1 + 5 \times 10^{-4}$. This simple estimation indicates that an intensity growth by a factor of 5 or 1000 upon a single reflection [4] seems unfeasible. Based on the consideration above, one can imagine that only after the accumulation of multiple reflections the effect of gain or loss becomes noticeable.

**III. Transmitted field in the low-index medium**

We consider a monochromatic infinite plane wave $E_{s0}^I$ with $s$-polarisation, incident from the high-index side at an angle of $\alpha$ from the normal to the plane. With the time dependence $\exp(\omega t)$ dropped as understood, the total field in the high-index medium is given by

$$E_y = E_{s0}^I \exp[-jk_0 n_h(x \sin\alpha + z \cos\alpha)] + E_s^R \exp[-jk_0 n_h(x \sin\alpha - z \cos\alpha)], \quad (16)$$

where $E_s^R = r_s E_{s0}^I$. If we write the field in the low-index medium

$$E_y = E^T_{s0} \exp[-jk_0 n_l(x \sin \beta + z \cos \beta)], \tag{17}$$

then the boundary conditions require that we have for the total field at $z = 0$:

$$E^T_s \exp[-jk_0 n_l(x \sin \beta)] = E^I_{s0} \exp[-jk_0 n_h(x \sin \alpha)] + r_s E^I_{s0} \exp[-jk_0 n_h(x \sin \alpha)]. \tag{18}$$

As we have seen in Sec. II, the amplitude change in the low-index medium is determined by the propagation length equal to the Goos-Hänchen shift and is thus a function of neither $x$ nor $z$ but the incident angle $\alpha$ through $\delta_s$ and is lumped into $E^T_s$. Notice that the finite propagation length in the gain or loss medium is essential to accommodate the steady-state solution. Noting Snell's law, we have

$$E^T_s = E^T_{s0} + \Delta E^T_s, \tag{19}$$

where

$$E^T_{s0} = 2n_h \cos \alpha / [n_h \cos \alpha - j(n_h^2 \sin^2 \alpha - n_l^2)^{1/2}] E^I_{s0} \tag{20}$$

and

$$\Delta E^T_s = \exp(-j\Phi_s)[\exp(-k_0 \kappa \delta_s) - 1] E^I_{s0}. \tag{21}$$

Here $E^T_{s0}$ is the transmitted field for the transparent case ($\kappa = 0$) and the second term $\Delta E^T_s$ describes the evanescent-field change as a result of atomic transitions induced by $E^T_{s0}$. This interpretation is supported by the result of a pioneering work by Carniglia *et al.* [25], in which it was experimentally shown that part of spontaneous radiation originated in the vicinity of the boundary excites the evanescent wave in the low-index medium and then is transformed into the outgoing homogeneous wave in the high-index medium.

Now we turn our attention to the problem of the Poynting vectors. There is no time-averaged power flow across the boundary for total reflection from a transparent low-index medium [18]. To see if the expression obtained above for the transmitted field gives an adequate description of the physical situation, we take the $z$-component of the Poynting vector. Including the time factor $\exp(j\omega t)$, we have for the $y$-component of the electric field

$$E_y = E^T_s \exp(-jk_0 n_h \sin\alpha\, x) \exp\{-k_0 n_l [(n_h/n_l)^2 \sin^2\alpha - 1]^{1/2} z\} \exp(j\omega t) \tag{22}$$

and, noting $n_l \gg \kappa$, for the $x$-component of the magnetic field

$$H_x = -jc(n_h^2 \sin^2\alpha - n_l^2)^{1/2} E^T_s \exp(-jk_0 n_h \sin\alpha\, x) \exp\{-k_0 n_l [(n_h/n_l)^2 \sin^2\alpha - 1]^{1/2} z\} \exp(j\omega t). \tag{23}$$

Here $c$ is the speed of light in vacuum. Taking the real parts of the electric and magnetic fields, (22) and (23), to calculate $S_z = -E_y H_x$, we find that, unlike the transparent case, the $z$-component of the time-averaged Poynting vector has a nonvanishing term proportional to $\exp(-k_0 \kappa \delta_s) - 1$, and that $S_z > 0$ for $\kappa < 0$ and $S_z < 0$ for $\kappa > 0$, as expected.

**IV. Summary**

A modification to Fresnel's theory of reflection has been formulated to account for evanescent gain or loss under the condition that the ratio of the imaginary to the real part of the index is sufficiently small. The derived expressions put forward a not-so-counterintuitive physical interpretation: as the atomic transitions responsible for gain or loss are induced by the evanescent field traveling along the boundary plane, the intensity change occurs in the direction parallel to the time-averaged energy flow in the low-index medium. It has been shown that the Poynting vector has a time-averaged component indicating there is a power flow across the boundary plane.

**Acknowledgements**

I am grateful for his generosity and scholarship to the late A. E. Siegman, who made available his then-unpublished manuscript on evanescent gain. Also, I owe a debt of gratitude to M. Fukuda, O. Karthaus, and F. Yasuda for their unfailing encouragement and support.

**Footnote**

[1] All the mode overlap factors are tabulated below. The expressions given in Problem 3.2, pp. 149-150 in [22] translate to the following in our notations. For TE modes, the fractional mode-medium overlaps for the cladding, the core, and the substrate are

$$\Gamma_s^l = d_s^l n_l^2 \cos^2\alpha (d_s^l + T + d_s^s)^{-1}(n_h^2 - n_s^2)^{-1},$$

$$\Gamma_s^c = T(d_s^l + T + d_s^s)^{-1} + d_s^l(n_h^2 - n_s^2 - n_l^2 \cos^2\alpha)(d_s^l + T + d_s^s)^{-1}(n_h^2 - n_s^2)^{-1} + d_s^s(n_h^2 - n_l^2 - n_s^2 \cos^2\alpha)(d_s^l + T + d_s^s)^{-1}(n_h^2 - n_l^2)^{-1},$$

and

$$\Gamma_s^s = d_s^s n_s^2 \cos^2\alpha (d_s^l + T + d_s^s)^{-1}(n_h^2 - n_l^2)^{-1},$$

respectively. These overlap factors are normalized so $\Gamma_s^l + \Gamma_s^c + \Gamma_s^s = 1$. For TM modes, we have

$$\Gamma_p^l = d_p^l n_l^2 \cos^2\alpha (d_p^l + T + d_p^s)^{-1}(n_h^2 - n_s^2)^{-1},$$

$$\Gamma_p^c = T(d_p^l + T + d_p^s)^{-1} + d_p^l(n_h^2 - n_s^2 - n_l^2 \cos^2\alpha)(d_p^l + T + d_p^s)^{-1}(n_h^2 - n_s^2)^{-1} + d_p^s(n_h^2 - n_l^2 - n_s^2 \cos^2\alpha)(d_p^l + T + d_p^s)^{-1}(n_h^2 - n_l^2)^{-1},$$

and

$$\Gamma_p^s = d_p^s n_s^2 \cos^2\alpha (d_p^l + T + d_p^s)^{-1}(n_h^2 - n_l^2)^{-1}.$$

Note also that $\Gamma_p^l + \Gamma_p^c + \Gamma_p^s = 1$. The fractions in the core, $\Gamma_s^c$ and $\Gamma_p^c$, are more commonly called the mode confinement factors.

**Figure caption**

Figure 1. Schematic illustration of a guided mode in a planar waveguide with a top cladding layer exhibiting gain or loss and transparent substrate. The symbol $\Delta L$ represents the distance between two successive reflections at the core-cladding boundary, $\Delta L_s$ for TE modes and $\Delta L_p$ for TM modes. The penetration depths $d^l$ in the cladding and $d^s$ in the substrate, respectively, are $d_s^l$ and $d_s^s$ for TE modes, and $d_p^l$ and $d_p^s$ for TM modes. Also, the Goos-Hänchen shift $\delta$ is $\delta_s$ for *s*-polarized waves and $\delta_p$ for *p*-polarized waves, respectively.

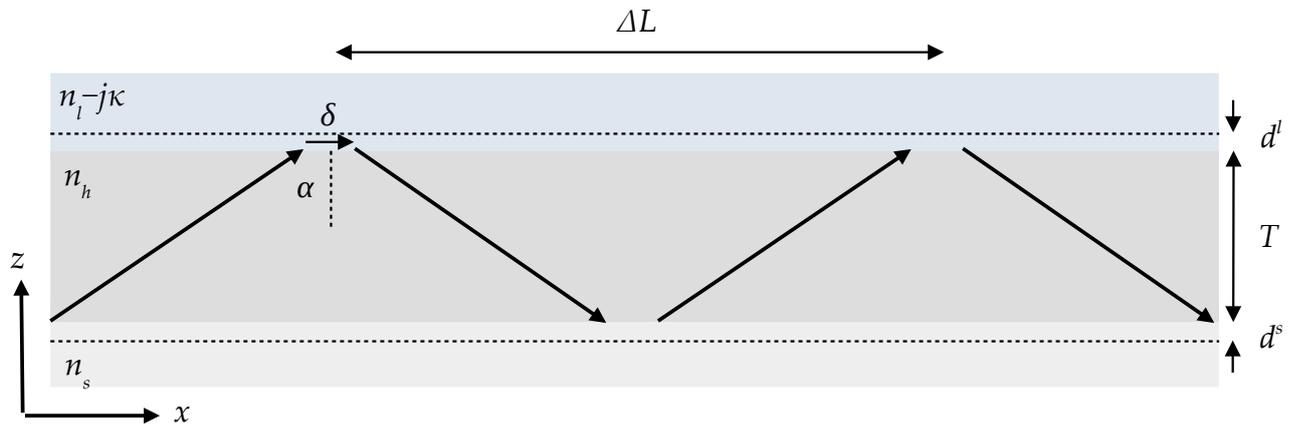

Figure 1      T. Kobayashi

17